\documentclass{ws-p10x7}
\def\as{\alpha_S}
 
\def\gg{\gamma^*\gamma^*}

\begin{document}
\title{
\hfill {\rm UR-1612}\\
\hfill {\rm ER/40685/950}\\
\hfill {\rm September 2000}\\
Recent Results on BFKL Physics}

\author{Lynne H.~Orr}

\address{Department of Physics and Astronomy, University of Rochester,\\
Rochester, NY 14627-0171, USA\\E-mail: orr@pas.rochester.edu}

\author{W.J.~Stirling}

\address{Departments of Physics and Mathematical Sciences, 
University of Durham\\ Durham DH1 3LE, England\\
E-mail: W.J.Stirling@durham.ac.uk}

\twocolumn[\maketitle\abstract{
Virtual photon scattering in $e^+e^-$ collisions can result in events with the 
electron-positron pair produced at large rapidity separation in
association with hadrons.  The BFKL equation resums large logarithms 
that dominate the 
cross section for this process. After a brief overview of analytic BFKL
resummation and its experimental status, we
report  on a Monte Carlo method for solving the BFKL equation that allows
kinematic  constraints to be taken into account.  We discuss results for 
$e^+e^-$  collisions using both fixed-order QCD and the BFKL
approach.  We conclude with some brief comments on the status of NLL
calculations.\\
\vskip 0.5\baselineskip
\noindent
Presented at 
XXXth International Conference on High 
Energy Physics, Osaka, Japan, July 27 -- August 2, 2000}] 

\section{Introduction }

Many processes in QCD can be described by a fixed order expansion in the 
strong coupling constant $\as$.  In some kinematic regimes, however,  each
power of  $\as$ gets multiplied by a large logarithm (of some ratio of relevant
scales),  and fixed-order calculations must give way to leading-log
calculations in which such terms are resummed.  The BFKL equation \cite{bfkl}
resums these large logarithms when they arise from multiple (real and virtual)
gluon emissions.  In the BFKL regime, the transverse momenta of the
contributing gluons are  comparable but they are strongly ordered in rapidity.

The BFKL equation can be solved analytically, and its solutions usually result in
(parton-level) cross sections that increase as the power $\lambda$, where 
$\lambda = 4C_A\ln 2\, \alpha_s/\pi \approx 0.5$.\footnote{$\lambda$ is also
known as $\alpha_P-1$.}
For example, in dijet production at large rapidity separation $\Delta$ in hadron 
colliders,\cite{muenav}
BFKL predicts for the parton-level cross section 
$\hat\sigma ~ e^{\lambda\Delta}$. In virtual photon scattering, for example in
 $e^+e^-$ collisions, where  the electron-positron pair at emerge with a large
rapidity separation and hadronic activity in between,\cite{brodskyetal} BFKL
predicts  $\sigma_{\gamma^*\gamma^*} ~ (W^2/Q^2)^\lambda$.  $W^2$ is the
invariant mass  of the  hadronic system (equivalently, the photon-photon
center-of-mass energy) and $Q^2$  is the invariant mass of either photon.

\section{Experimental Status and Improved Predictions}
The experimental status of BFKL is ambiguous at best, with existing results 
being far from definitive.  The data tend to lie between the predictions of
fixed-order QCD and analytic solutions to the BFKL equation.   This happens,
for example, for the azimuthal decorrelation in dijet production at the Fermilab
Tevatron\cite{dzeroazi} and for the virtual photon cross section at
LEP.\cite{maneesh} Similar results are found in $ep$ collissions at
HERA.\footnote{One exception   is the ratio of the dijet production cross
sections at center of mass energies 630 GeV and 1800 GeV at the Tevatron,
where the measured ratio lies above  {\it all}  predictions.\cite{pope}}

It is not so surprising that analytic BFKL predicts stronger effects 
than seen in data.  Analytic BFKL solutions implicitly contain sums over 
arbitrary numbers of gluons with arbitrary energies, but the kinematics are 
leading-order only.  As a result there is no kinematic cost to emit gluons, and 
energy and momentum are not conserved, and BFKL effects are
artifically enhanced.

This situation can be remedied by a Monte Carlo implementation of solutions to
the BFKL equation.\cite{os,schmidt}  In such an implementation the BFKL
equation is solved by iteration, making  the sum over gluons explicit.  Then
kinematic constraints can be implemented  directly, and conservation of energy
and momentum is restored.  This tends to lead to  a suppression of BFKL-type
effects.  The Monte Carlo approach has been applied to dijet
production at hadron colliders,\cite{os,schmidt,osmore} leading to 
better (though still not perfect) agreement with the dijet azimuthal
decorrelation data.\cite{os}  Applications to forward jet production at
HERA and to virtual photon scattering in $e^+e^-$ collisions are underway;
an update on the latter appears in the next section.

\section{$\gg$ Scattering:  A Closer Look}

BFKL effects can arise in $e^+e^-$ collisions via the scattering 
of virtual photons emitted from the initial $e^+$  and $e^-$.
The scattered electron and positron appear in the forward and backward
regions (``double-tagged'' events)
with hadrons in between.  With total center-of-mass energy $s$, 
photon virtuality $-Q^2$, and photon-photon invariant mass ($=$ invariant
mass of the final hadronic system) $W^2$, BFKL effects are expected 
in the kinematic regime where $W^2$
is large and $$s>>Q^2>>\Lambda_{QCD}^2.$$  
At fixed order in QCD, the dominant process is four-quark production with
$t$-channel gluon exchange (each photon couples to a quark box; the quark
boxes are connected via the gluon).  The corresponding BFKL
contribution arises from diagrams with a gluon ladder attached to the 
$t$-channel gluon.

\begin{figure}
\epsfxsize200pt
\figurebox{}{}{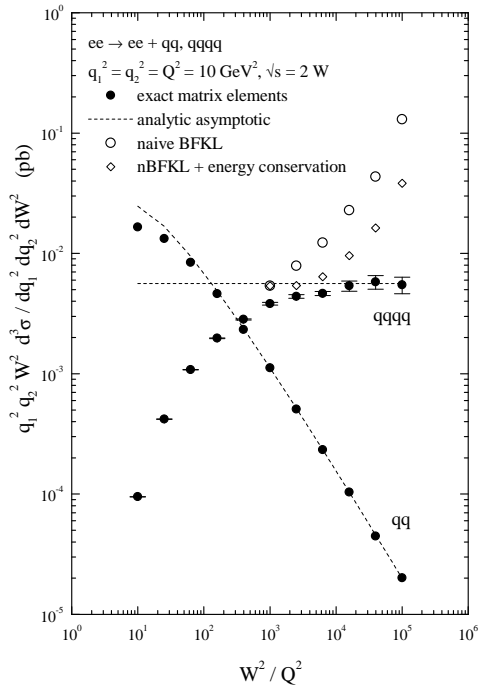}
\caption{Exact (closed data points) and analytic asymptotic (dashed line) 
$e^+e^- \to e^+e^-q \bar q$  and 
$e^+e^- \to e^+e^-q \bar q q \bar q$  cross sections versus 
$W^2/Q^2$ at fixed $W^2/s = 1/4$.  Also shown:  analytic BFKL without (open 
circles) and with (open diamonds) energy conservation imposed.} 
\label{fig:compare}
\end{figure}

The relative contributions of fixed-order QCD and BFKL are most easily
understood by looking at  
$$
W^2 Q_1^2 Q_2^2 \; {d^3 \sigma \over  d W^2 d Q_1^2 d Q_2^2 } 
$$
as a function of $W^2/Q^2$ for fixed $\sqrt{s}/W$.  The asymptotic regime then
corresponds to large $W^2/Q^2$.  This quantity is shown in Figure 1 for
$Q_1^2=Q_2^2=Q^2=10\ {\rm GeV}^2$ and $\sqrt{s}=2W$.  The solid points are
the QCD calculations of two-quark (`qq') and four-quark production (`qqqq'); we
see that  the latter dominates for large $W^2/Q^2$ and approaches a
constant asymptotic  value.  In contrast, the analytic BFKL result, shown with
open circles,  rises well above that of fixed-order QCD.  The diamonds show
analytic BFKL with energy conservation imposed, but not exact kinematics;
it can be interpreted as an upper limit for the Monte Carlo prediction,
which is in progress.

It is important to note in Figure 1 that although BFKL makes a definite
leading-order prediction for the behavior of the cross section as  
a function of $W^2/Q^2$, the origin  in
$W^2/Q^2$  (i.e., where  BFKL 
meets asymptotic QCD) is  {\it not} determined in leading order.
  We have chosen $W^2/Q^2=10^3\ {\rm GeV}^2$ as a 
reasonable value where the QCD behavior is sufficiently asymptotic for 
BFKL to become relevant, but another choice might be just as reasonable.  
Only when higher order corrections are computed can the BFKL prediction
be considered unique.   

\begin{figure}
\vspace*{-1.5cm}
\epsfxsize200pt
\figurebox{}{}{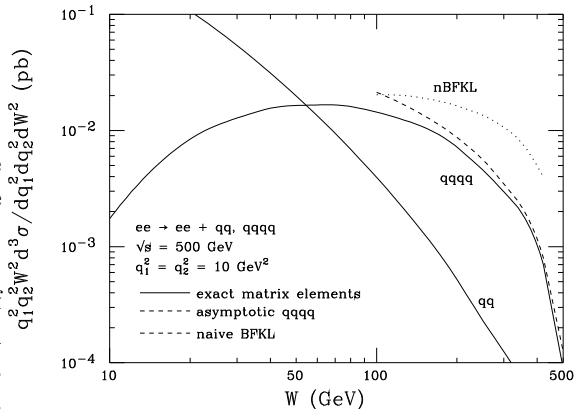}
\vspace*{-2.5cm}
\caption{Exact (solid lines) and analytic asymptotic (dashed line) 
$e^+e^- \to e^+e^-q \bar q$  and 
$e^+e^- \to e^+e^-q \bar q q \bar q$  cross sections versus 
$W^2/Q^2$ at fixed $\sqrt{s}=500\ {\rm GeV}$.  Also shown:  analytic BFKL
(dotted line).}  \label{fig:lc}
\end{figure}

From an experimental point of view, the cross
section at fixed $\sqrt{s}$ is more directly relevant.  Figure 2 shows 
$W^2 Q_1^2 Q_2^2 \; {d^3 \sigma \over  d W^2 d Q_1^2 d Q_2^2 }$
for a linear collider energy $\sqrt{s}=500\ {\rm GeV}$.  The solid lines
show the exact fixed-order QCD prediction.  The dashed line
is the asymptotic four-quark production cross section, and the 
dotted line is the analytic BFKL prediction. Now we see that all
of the curves fall off at large $W$, but the BFKL cross section lies well
above the others.

\subsection{$\gg$ Scattering at LEP}

The L3 collaboration at LEP have measured  the $\gg$ cross section by
dividing the double-tagged $e^+e^-$ cross section by the $\gg$ 
luminosity.\cite{maneesh}  They present their results, 
for $\sqrt{s}=183$ and $189$--$202$ GeV, with 
$Q^2=14\ {\rm GeV}^2$ and $Q^2=15\ {\rm GeV}^2$, respectively, as a function
of $y=\ln(W^2/Q^2)$.  In this variable the asymptotic QCD four-quark cross
section is flat, and the analytic BFKL cross section rises, similarly to 
Figure 1.  The data lie between the two predictions, and the higher
statistics data at the higher energies show a clear rise in the data, though
not as steep as predicted by analytic BFKL.  

We expect that the BFKL Monte Carlo prediction (in progress) will be closer to
the data.  But one can also
ask whether the asymptotic QCD limit for four-quark production 
is appropriate here.  We compare the exact and asymptotic QCD curves
at the LEP energy $\sqrt{s}=183\ {\rm GeV}$  in Figure 3 (note that this is
the undivided $e^+e^-$ cross section that includes the photon luminosity).
The values of $W$ corresponding to the LEP measurements range between 
about 15 and 90 GeV.  Comparing   four-quark
predictions, we see that the exact curve is not close enough to the 
asymptotic in this region for the asymptotic QCD limit to
be appropriate.  Furthermore, the ratio of exact to asymptotic
results --- which is proportional to the $\gg$ cross section --- 
{\it rises} in this region.  The QCD prediction is not flat at 
all.\footnote{This does not automatically imply that fixed-order QCD describes
the data, because there are unresolved normalization issues involved.}  
Until 
the fixed-order QCD and BFKL Monte Carlo predictions are sorted out,
it is not clear what we can  conclude from the data.

\begin{figure}
\vspace*{-1.5cm}
\epsfxsize190pt
\figurebox{}{}{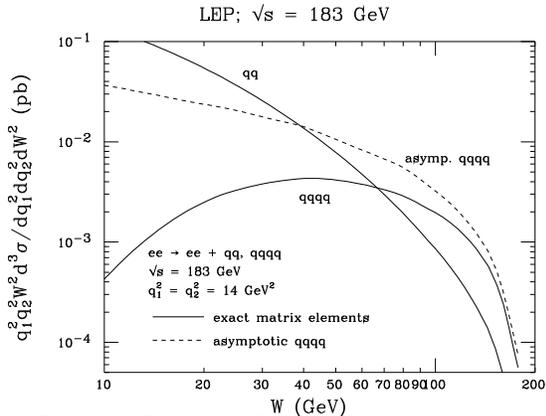}
\vspace*{-2.5cm}
\caption{Exact (solid lines) and analytic asymptotic (dashed line) 
$e^+e^- \to e^+e^-q \bar q$  and 
$e^+e^- \to e^+e^-q \bar q q \bar q$  cross sections versus 
$W^2/Q^2$ at fixed $\sqrt{s}=183\ {\rm GeV}$.}  \label{fig:lep}
\end{figure}

\section{Status of NLL Corrections}

It is apparent that, although it is not yet clear 
whether BFKL is necessary to describe the data in hand, leading-order analytic
BFKL is  not sufficient. We need BFKL at next-to-leading order (strictly
speaking, next-to-leading log order).  This has been accomplished
after 10 years of heroic efforts by Fadin, Lipatov and many others
(see \cite{salam} for a review, complete  references, and more details
about what follows).  The bad news is that the solutions appear to
be large, unstable, and capable of giving negative cross sections.
The good news is that there is much progress in understanding these 
problems, which can mostly be traced to the fact that at NLL gluons
can be close together in rapidity, leading to collinear
divergences.  Several methods for solving this problem are summarized in
\cite{salam}, and  the NLL BFKL corrections appear to 
be coming under control.

\section{Conclusions}

In summary, BFKL physics is a complicated business.  Tests are being performed
in a variety of present experiments (Tevatron, HERA, LEP) and there is 
potential for the future as well (LHC, LC).  Unfortunately, comparisons between
theory and experiment are not straightforward; leading order BFKL is
insufficient, and subleading corrections such as kinematic constraints can be
very important.  A worst-case scenario which is not ruled out may be that 
we cannot reach sufficiently asymptotic regions in experiments to see 
unambiguous BFKL effects.  However, reports of the demise of BFKL physics 
due to instability of the next-to-leading-order corrections are greatly
exaggerated, and the source of the large corrections is understood and they
are being brought under control.  In summary, the jury on BFKL physics 
is still out, but there continues to be much progress. 

\section{Acknowledgments}
\noindent
Work supported in part by the U.S. Department of Energy and the U.S. 
National Science Foundation,
under grants DE-FG02-91ER40685 and  PHY-9600155.

\end{document}